\documentclass[prb,twocolumn,showpacs,preprintnumbers,amsmath,amssymb,superscriptaddress]{revtex4}
\usepackage{epsf}
\usepackage{graphicx}
\usepackage{bm}
\usepackage{amsmath}
\usepackage{color}
\usepackage{notes2bib}
\usepackage{epstopdf}
\usepackage{lineno}
\usepackage{float}

\bibnotesetup{
note-name = , use-sort-key = false }
\begin{document}

\title{Topological Hall effect for electron scattering on skyrmions in external magnetic field}
\author{K.~S.~Denisov}
\email{denisokonstantin@gmail.com}
\affiliation{Ioffe Institute, St.Petersburg, 194021 Russia}
\affiliation{Lappeenranta University of Technology, FI-53851 Lappeenranta, Finland}
\author{I.~V.~Rozhansky}
\affiliation{Ioffe Institute,
St.Petersburg, 194021 Russia} \affiliation{Lappeenranta University of
Technology, FI-53851 Lappeenranta, Finland}
\author{M.~N.~Potkina}
\affiliation{SPbGU, St.Petersburg, 198504 Russia}
\affiliation{University of Iceland, VR-III, Reykjavik, Iceland}
\author{I.~S.~Lobanov}
\affiliation{ITMO University, St.Petersburg, 197101 Russia}
\author{E.~L\"ahderanta}
\affiliation{Lappeenranta University of Technology, FI-53851 Lappeenranta, Finland}
\author{V.~M.~Uzdin}
\affiliation{SPbGU, St.Petersburg, 198504 Russia}
\affiliation{ITMO University, St.Petersburg, 197101 Russia}

\begin{abstract}
We consider topological Hall effect (THE) in thin ferromagnetic films due to electron scattering on magnetic skyrmions in the presence of the relatively strong external magnetic field.
We account for the effect of the magnetic field on a skyrmion structure and describe the hallmarks of THE differentiating it from ordinary and anomalous Hall effects. 
We have found that, although in typical ferromagnets the variation of magnetic field in the range  1-5 T substantially affects the skyrmion size, THE changes rather weakly 
 remaining quite robust.
 Therefore, the magnitude of THE is primarily determined by the skyrmions sheet density $n_{sk}$, 
 being comparable to the magnitude of the ordinary Hall effect (OHE) at $n_{sk}=10^{11}$ cm$^{-2}$.
The sign of THE is opposite to that of OHE for skyrmions with positive vorticity,
while for antiskyrmions the signs are the same. 

\end{abstract}


\date{\today}

\maketitle

\section{Introduction}

Chiral magnetic quasi-particles such as magnetic skyrmions have recently gained a considerable interest~\cite{NagaosaNature,Romming,BraunNano,volovik1987linear}. 
These vortex-like magnetic textures characterized by a nonzero topological charge can be stable above room temperature \cite{Woo16} and are able to move very fast by electric current, which makes them considered for
using in dense and fast racetrack memory\cite{racetrack,fert2013skyrmions,muller2017magnetic,zhang2015magnetic}. 
Magnetic skyrmions affect conductivity of the system 
leading to the topological Hall effect (THE)\cite{Muhl_MnSi_Science,Machida_PRL,FeGe_THE,DiscretHall,THEAphase,AronzonRozh,Nagaosa2018}. 
The appearance of this additional contribution to the Hall response is due to the
electron exchange interaction with 
 non-collinear magnetization textures~\cite{Ye1999,Tatara,Lyana-Geller1,BrunoDugaev,ourSciRep2017}. 
While this effect is very attractive as it provides the possibility to readout the racetrack memory state by electrical means,  
it is still challenging to differentiate the topological contribution from ordinary and anomalous Hall effects~\cite{maccariello2018electrical,ahadi2017evidence,DiscretHall}.
As has been shown theoretically, THE has a  
nontrivial 
dependence both on the mesoscopic disorder~\cite{nakazawa2017topological,Arab_Papa} and on the skyrmion structure~\cite{ourSciRep2017}.
%
Although in ferromagnets the latter is strongly affected by the external magnetic field $B$, the dependence of THE on $B$ has not been thoroughly studied so far. 

In this paper we consider THE  due to electron coherent scattering on individual magnetic skyrmions 
in the presence of 
external magnetic field $B$.
We describe 
the 
hallmarks of THE, which distinguish it from ordinary and anomalous Hall effects,
taking into account the effect of the magnetic field on the  skyrmion structure.
Our results 
are applicable to 
ferromagnetic thin films with disordered arrays of quasi-2D magnetic skyrmions and  2D  electrons 
filling a size quantization subband. 
We analyze the features of THE in the diffusive transport regime paying attention to the qualitative difference with previously studied mesoscopic theory~\cite{BrunoDugaev,Arab_Papa}. 

The paper is organized as follows: in section \ref{Sec2} we review the properties of asymmetric electron scattering on individual magnetic skyrmions, in section \ref{Sec3}  the formation and the inner structure of a magnetic skyrmion under external magnetic  field is discussed, in \ref{Sec4} we focus on THE in the disordered ensemble of skyrmions in external magnetic field.

\section{Scattering on a single magnetic skyrmion}
\label{Sec2}

\begin{figure*}[t]
	\centering	
	\includegraphics[width=\textwidth]{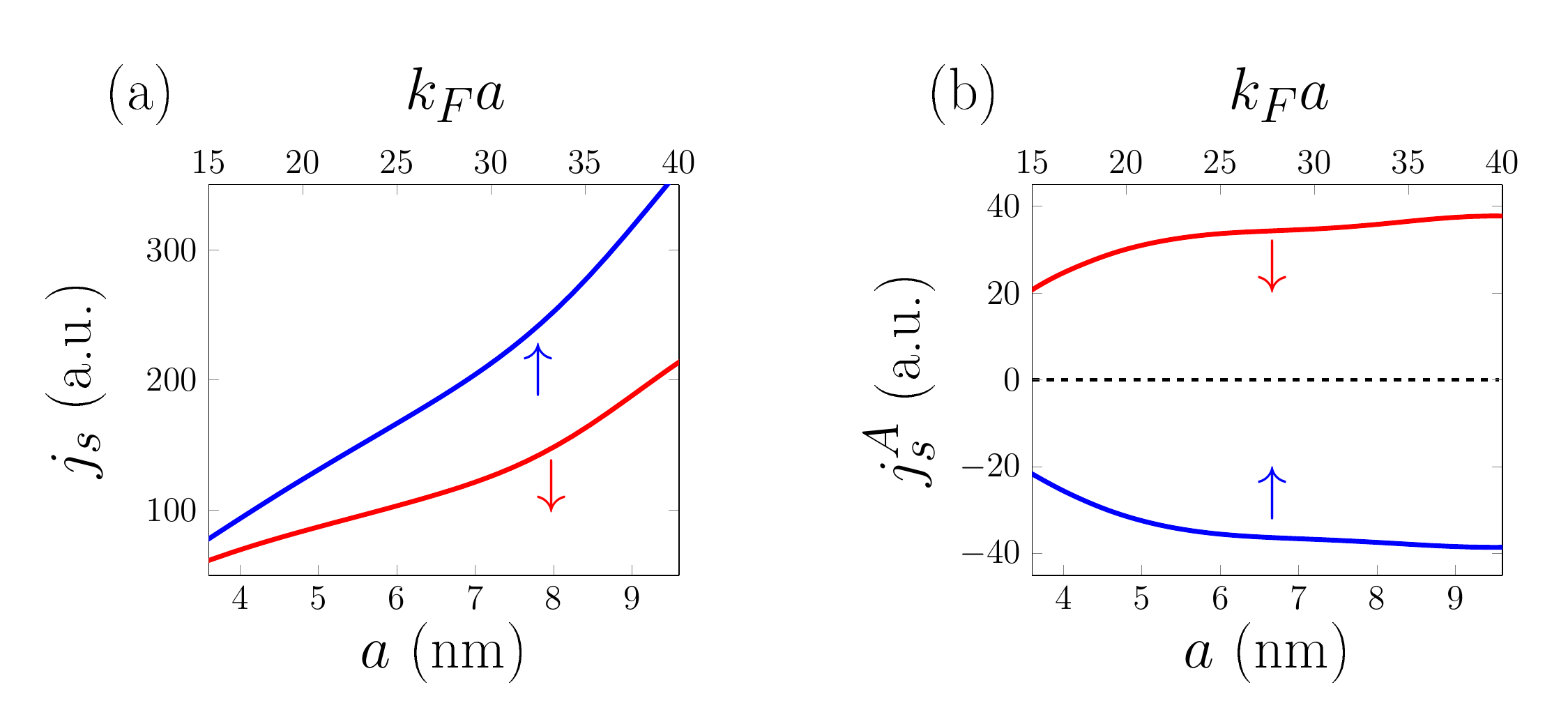}
	\caption{The dependence of total symmetric $j_s$ (a) and asymmetric $j_s^A$ (b) scattering rates on a skyrmion diameter $a$ and the adiabatic parameter $\lambda_a$. In the adiabatic regime $\lambda_a \gg 1$ spin-up and spin-down electrons are scattered in opposite transverse directions. The parameters are: $E_F = 1$ eV, $\Delta =1$ eV, $m_{\ast}=0.6 m_0$ , $m=+1$, $\Theta(r)$ is given by (\ref{appr}) with fitting parameters discussed in Sec.\ref{Sec3}.}
	\label{Fig1}
\end{figure*}

Let us start with the problem of 2D  scattering of an electron on an individual magnetic skyrmion~\cite{ourSciRep2017}. {We
assume that the electrons fill  one 2D size quantization subband of a thin  
ferromagnetic film and, therefore, characterized by in-plane wavevector $k$ and spin projection $s$ on $z$ axis perpendicular to the film plane.}
The electron exchange interaction with magnetization field described by a spatial profile $\boldsymbol{S}(\boldsymbol{r})$ is given by the Hamiltonian:
\begin{equation}
\label{eq_Hsk}
H_{ex} = -\frac{\Delta}{2} \boldsymbol{S} ({\boldsymbol{r}}) \cdot \boldsymbol{\sigma},
\end{equation}
where $\Delta$ is the exchange splitting, $\boldsymbol{\sigma}$ is the vector of Pauli matrices 
and $\boldsymbol{r}=(r,\varphi)$ is a planar radius vector with magnitude $r$, and polar angle $\varphi$. 
The skyrmion is described by $\boldsymbol{S}(\boldsymbol{r}) = (\sin{\Theta} \cos{\Phi} , \sin{\Theta} \sin{\Phi}, \cos{\Theta})$; the function $\Theta(r)$ gives the radial profile, $\Phi(\varphi) = m \varphi + \gamma$ describes the in-plane rotation of magnetization; $\gamma$ is helicity and $m$ is vorticity. For a skyrmion  $m=+1$, while  $m=-1$ corresponds to  antiskyrmion~\cite{Parkin}. 

Outside of the skyrmion 
the magnetization field is uniform with $S_z=1$.
The perturbation of the background magnetization 
{in a skyrmion}
causes 
the electron scattering on the potential:
\begin{equation}
    \label{eq_Psk}
    V_{sk} = - \frac{\Delta}{2}
    \begin{pmatrix}
     \cos{\Theta}-1 & e^{- i m \varphi - i\gamma} \sin{\Theta}
     \\
      e^{i m \varphi + i\gamma} \sin{\Theta}  & 1-\cos{\Theta}
    \end{pmatrix}
\end{equation}
The dependence of $V_{sk}$ on polar angle $\varphi$ 
leads to the appearance of asymmetric part in the electron differential scattering cross-section:
\begin{equation}
    \label{eq_cross1}
    \frac{d \sigma_{s's}}{d \varphi} = G_{s's}(\varphi) + \Sigma_{s's}(\varphi),
\end{equation}
where $s,s'$ {denote spin projection on $z$ axis normal to the film plane},
$G_{s's}(\varphi) = G_{s's}(-\varphi)$ is symmetric with regard to the scattering   angle $\varphi$, and $\Sigma_{s's}(\varphi) = - \Sigma_{s's}(-\varphi)$ is the asymmetric term responsible for THE. 
 
The properties of $\Sigma_{s's}$ appear to be qualitatively different depending on the adiabatic parameter $\lambda_a=(\Delta/{2E_F})\left(k_Fa\right)$, where $a$ is the skyrmion size, $E_F$ is the Fermi energy of the electrons\cite{ourSciRep2017}, 
$\hbar k_F = (2m_{\ast} E_F)^{1/2}$, $m_\ast$ is the electron effective mass. In this paper we focus on the adiabatic regime $\lambda_a \gg 1$ typical for strong ferromagnets, when the exchange interaction is strong and spin-flip scattering channels are suppressed.


 
 



As a model system we consider 
a thin film of FePd on Ir(111) surface~\cite{Hagemeister}. 
Electrons in the few-monolayer-thick film experience size quantization 
with their wavevector parallel to the surface of the sample~\cite{HOka}. 
Skyrmions in such films are strongly affected by the external magnetic field: {the increase of $B$ directed opposite to magnetic moment in the skyrmion core reduces the skyrmion size}~\cite{Bessarab} and, 
consequently, affects the magnitude of THE.

In order to investigate 
this effect, let us consider the dependence of the scattering cross-section 
on the skyrmion size $a$. We calculate the cross-section (\ref{eq_cross1}) using the phase function method~\cite{ourSciRep2017}. 
To analyze the Hall conductivity
we introduce dimensionless total symmetric $j_s$ and asymmetric $j_s^A$ rates as
\begin{equation}
j_s = 2\pi k_s \int G_{ss} d\varphi
\hspace{0.5cm}
\displaystyle j_s^A =  2\pi k_s \int \Sigma_{ss} \sin{\varphi} d\varphi,
\end{equation}
where $\hbar k_s = (2m_{\ast} (E_F + s \Delta))^{1/2}$ is the spin-dependent wavevector, {$s$ denotes the spin projection on $z$-axis}.

Fig.~\ref{Fig1} shows the dependence of $j_s$ and $j_s^A$ on the skyrmion size $a$ calculated 
using the parameters: $E_F = 1$ eV, $\Delta = 1$ eV, $m_{\ast} = 0.6 m_0$, the lattice constant $a_0 = 0.27$ nm. The skyrmion radial profile is {taken as}:
\begin{equation}
\Theta(r)=\pi+\sum_{{t=\pm 1}}\arcsin\left(\tanh\left(\frac{{-r +t c}}{{w/2}}\right)\right).
\label{appr}
\end{equation}
as was proposed in Ref.~\cite{Romming}. 
The parameters $w,c$ are discussed in section \ref{Sec3}. 
As can be clearly seen from Fig.\ref{Fig1}, the total $j_s$ increases with the skyrmion size. 
This is quite expected because for 2D scattering on a cylinder the cross section is linearly proportional to its radius.
The asymmetric part of the cross-section which determines THE also increases with the skyrmion size.
It has the opposite sign for spin up and spin down electrons as expected in the adiabatic regime~\cite{ourSciRep2017}. However, our calculations show that for large skyrmion size $j_s^A$ saturates. The physical reason for this saturation is that 
the increase of the scatterer size is compensated by the decrease of the
magnetization gradients.



\section{Skyrmions in the external magnetic field}
\label{Sec3}

Topological charge of a skyrmion is an  integer number, which cannot be changed under continuous variation of magnetization~\cite{NagaosaNature}. The difference between
the topologically charged 
skyrmion and ferromagnetic state with $Q$=0 leads to stability of a skyrmion in relation to thermal fluctuations and random potential disorder 
in continuous limit. On a discrete lattice, topological arguments, strictly speaking, are inapplicable and one can only talk about the energy barriers between states and the pre-exponential factor in the Arrhenius law for the magnetic state lifetime~\cite{Physica_b,JMMM,Bessarab}. The lifetime and, therefore, the stability of a skyrmion,  strongly depends on its size. 

Properties of the skyrmion including its size $a$ are determined by the intrinsic parameters of the material as well as by external magnetic field. 

We describe the skyrmion on 2D triangular lattice by Heisenberg-like Hamiltonian:

\begin{equation}\label{eq:heisenberg}
\begin{gathered}
E[{\bf S}]=-\sum_{<j,k>}(J_{jk} {\bf S_j\cdot S_k+D_{j,k}\cdot(S_j\times S_k))}\\
-K \sum_j {S^2_{j,z}}- \mu {\bf B} \sum_j  \bf S_j,
\end{gathered}
\end{equation}
 where $ \bf S_j $ is the unit vector along the magnetic moment on site {\it j}, $\mu$ is the 
 magnitude of the magnetic moments, which is assumed to be the same for all sites, $\it J_{jk}$ is the exchange parameter, which we assume to be non-zero only for the nearest-neighbour sites, $\bf D_{jk}$ is the Dzyaloshinskii-Moriya vector and {\it K} is the anisotropy parameter. $\bf D_{jk}$ is chosen to lie in the plane of the lattice perpendicular to the vector connecting atomic sites j and k. Vector of anisotropy points perpendicular to the lattice plane.


Parameters were chosen to describe the skyrmion states in PdFeIr(111): K=0.07J, D=0.32J, J=7 meV, $\mu=3\mu_B$~\cite{JMMM}. Without magnetic field 
a spin spiral structure is formed in this system, but in a magnetic field of $B\sim2-5$ T a single skyrmion state is stable at a temperature of 8K and below~\cite{Hagemeister}.
At a high magnetic field the lifetime of skyrmion state decreases and needs lower temperature to be observed~\cite{JMMM}.

\begin{figure}
	\centering	
	\includegraphics[width=0.5\textwidth]{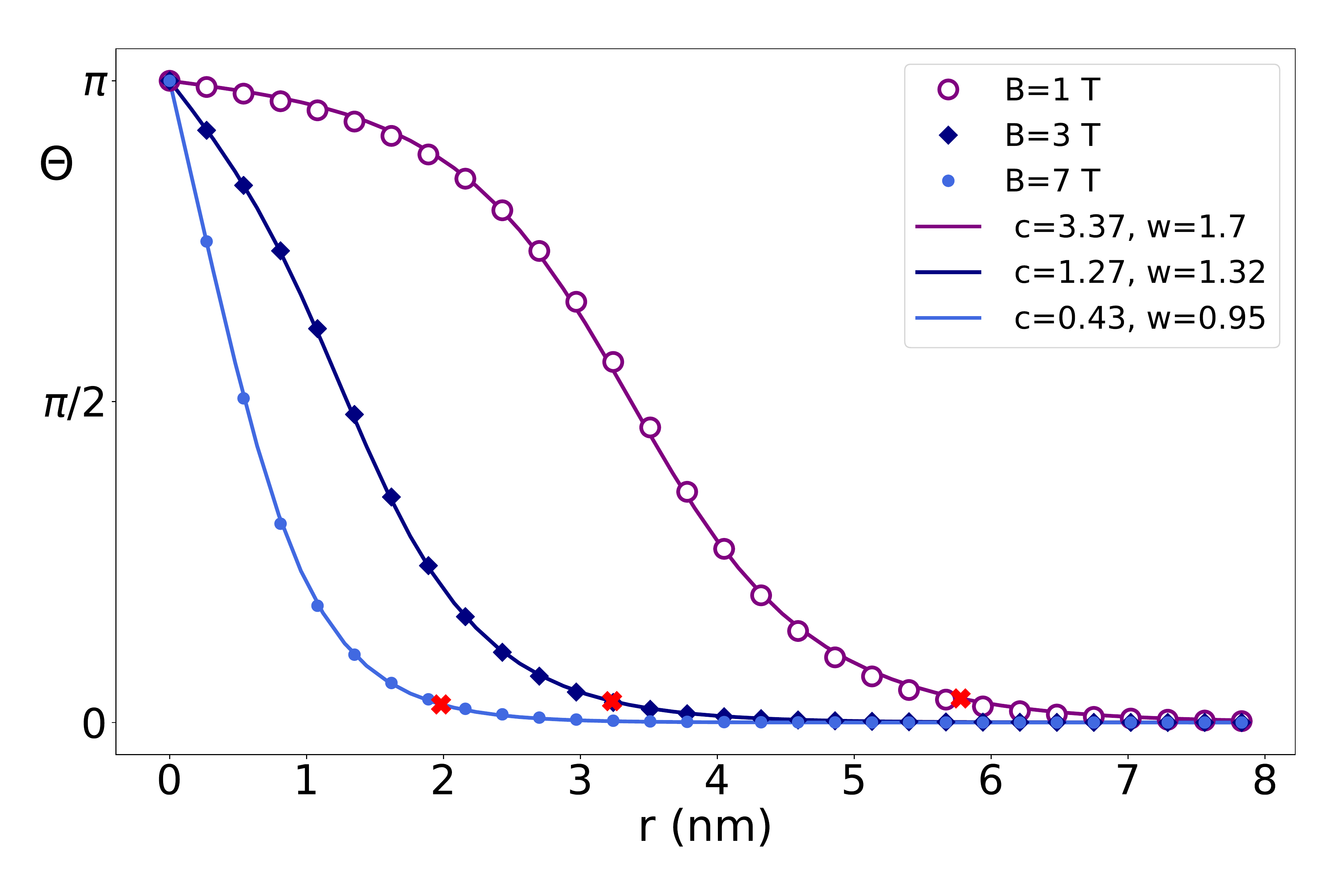}
	\caption{Profile $\Theta(r)$ of skyrmion at magnetic field B=1, 3 and 7 T and their approximation by function (\ref{appr}). Red labels denote radius of skyrmion.}
	\label{profile}
\end{figure}

Fig.~\ref{profile} shows calculated skyrmion profiles $\Theta$ for several values of the 
magnetic field and their approximation by the function $\Theta(r)$ from Eq.~(\ref{appr}). 
Diameter of the skyrmion is determined as 
diameter of a circle with $\Theta(r)=0.1$ as shown by red labels in Fig.~\ref{profile}.

\begin{figure}[H]
	\centering	
	\includegraphics[width=0.47\textwidth]{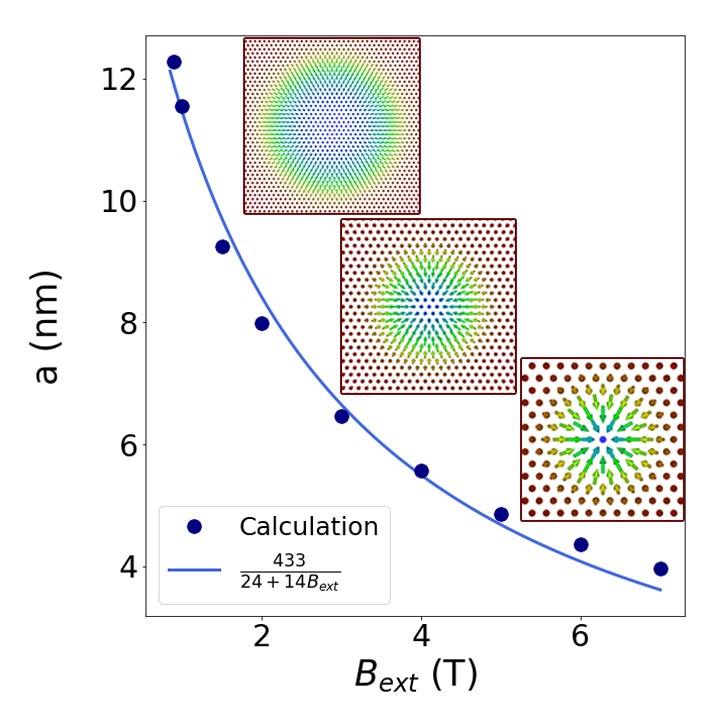}
	\caption{Dependence of skyrmion diameter $a$ on magnetic field $B_{ext}$. Insets show skyrmion configurations at magnetic field B=1, 3 and 7 T from top to bottom. }
	\label{FigRB}
\end{figure}

Dependence of the skyrmion radius on the parameters of the model was estimated previously\cite{zhang2015magnetic} with the expression 
$R_{sk}={D\pi^3}/\left({2K\pi^2+8\mu B}\right)$ with micromagnetic parameters $D, K$ using linear ansatz for skyrmion profile~\cite{Bogdanov1}. As our calculations demonstrate, this formula cannot reproduce radius for arbitrary  values of parameters D, K, J and B. However, it can be used to approximate the dependence of the radius on the magnetic field if the other parameters are considered as fitting values. Fig.~\ref{FigRB} shows this dependence, the insets illustrate the skyrmion configurations at 3 particular values of the magnetic field B=1, 3, 7 T.
Profiles (\ref{appr}) calculated for various magnetic fields were further used for the calculations of THE for scattering on a single skyrmion. 



\section{Transport in disordered skyrmionic system}
\label{Sec4}
\begin{figure*}[t]
	\centering	
	\includegraphics[width=1\textwidth]{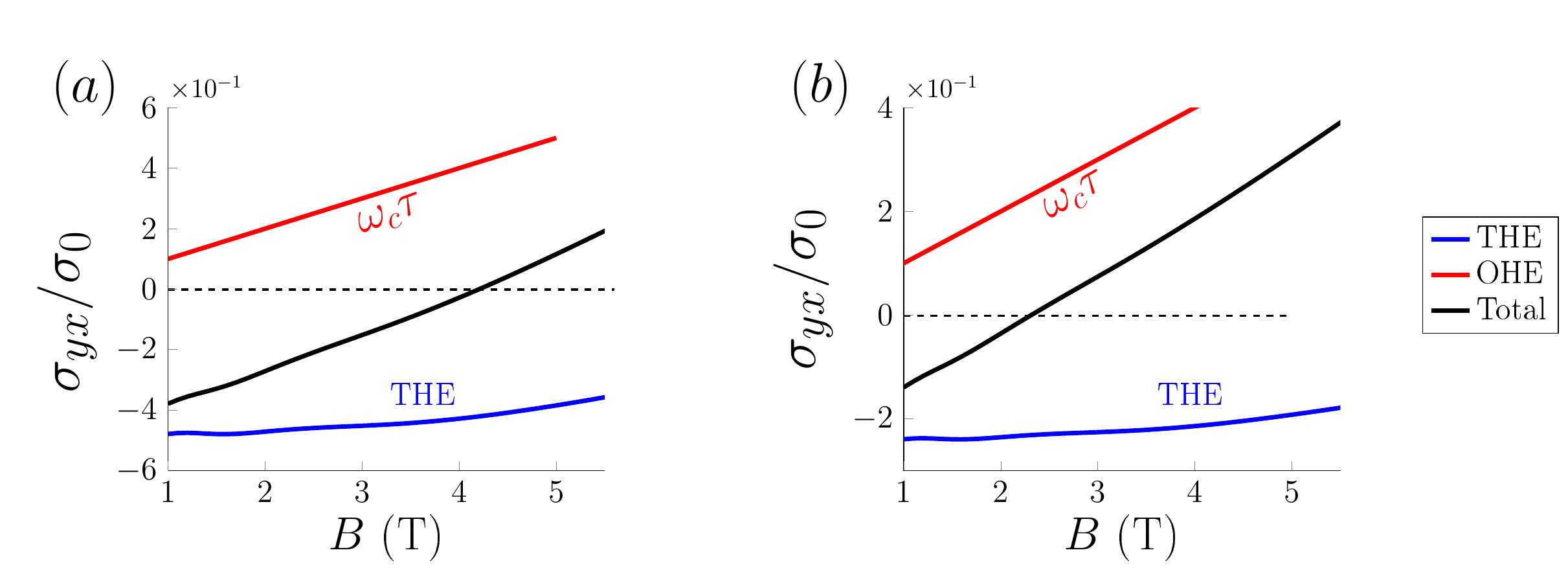}
	\caption{The dependence of topological $\sigma_{yx}^T$,  ordinary $\sigma_{yx}^{O}$ and total $\sigma_{yx}^T + \sigma_{yx}^{O}$ Hall conductivity on the external magnetic field $B$ for 2 skyrmion sheet concentrations $n_1 = 2.5\times 10^{11}$ cm$^{-2}$ (a), $n_2 = 1.25\times 10^{11}$ cm$^{-2}$ (b).}
	\label{FigHall}
\end{figure*}

In this section we describe the general features of THE in disordered array of skyrmions in a relatively strong magnetic field $B = 1-5$ T. We show that 
for a typical skyrmion sheet density $n_{sk} \ge 10^{11}$ cm$^{-2}$ 
the magnitude of THE is comparable or even  exceeds that of the 
ordinary Hall effect (OHE).  The sign of THE appears to be opposite to the sign of OHE for a skyrmion ($m=1$) configuration and coicindes with that of the OHE for antiskyrmion ($m=-1$). 

The properties of THE in a magnetic film with
skyrmions depend on the transport regime. 
 We assume that the  free electrons 
 localized in the conducting film are scattering on skyrmions as well as on non-magnetic impurities. The latter scattering gives the dominating contribution to the transport time and determine the  mean-free path $l$. 
In the mesoscopic regime valid for dense skyrmion arrays and skyrmion crystals 
the distance $d_{sk}$ between the skyrmions centers is of the same order as their size $a$, while the mean free path $l \gg d_{sk},a$.
If the electron wavefunction remains coherent over a few skyrmions so that 
$l_{\phi} \gg d_{sk}$,
where 
$l_{\phi}$ is the phase coherence length,
the scattering can be described in the framework of the meanfield approach~\cite{BrunoDugaev}. 
In this case an effective uniform magnetic field is determined by the global topological properties of the magnetization field: $B_t \sim \Phi_0 /s_0$, where $\Phi = h c/ |e|$ is the magnetic flux quantum and $s_0 \sim d_{sk}^2 \sim a^2$ is an
area per one skyrmion. 
In the mesoscopic regime an increase of the skyrmions size 
implies the increase of $d_{sk}$ 
leading to the overall decrease of THE magnitude.



Here we consider the transport in the diffusive regime when the conditions $l \gg a$, $a<l_{\varphi} \ll d_{sk}$ are fulfilled. 
In this case the electron interaction with the skyrmions should be considered as scattering on individual skyrmions~\cite{ourSciRep2017}. 
At that, the skyrmion size $a$ and skyrmion sheet density $n_{sk}$ affect the magnitude of THE independently. 
According to Fig.~\ref{Fig1} the increase of $a$ 
leads to the increase  of 
the transverse scattering rate and, therefore, to the increase of THE. 
While the total scattering cross-section linearly increases with the skyrmion size, the spatial gradients of magnetization  
field forming
the skyrmion texture and underlying the asymmetric scattering decrease. 
Interestingly, this leads to the saturation of the  asymmetric part of the cross-section as shown in Fig.~\ref{Fig1}.
As for the skyrmions sheet density $n_{sk}=d_{sk}^{-2}$, the magnitude of THE depends linearly on $n_{sk}$ as one would naturally expect from Drude theory. The skyrmions sheet density, in principle, 
can depend on the  magnetic field as 
the latter affects the skyrmion lifetime. In our analysis we do not account for this effect, i.e. we assume 
that  the temperature is low enough so that the skyrmion lifetime always exceeds the time of a measurement.

Let us consider relatively high external magnetic fields so that the background magnetization is saturated. Then the contribution to the transverse conductivity  from the anomalous Hall effect (AHE) does not depend on the external magnetic field and can be
identified as the constant contribution to the transverse current. 
The only two magnetic field dependent contributions come from OHE and THE. Let us compare them in more detail. 

For OHE the transverse conductivity is given by
\begin{equation}
\label{eqNormHall}
\sigma_{yx}^{O} = \sigma_0\left(\omega_c \tau\right),\,\,\,\,\,\,\sigma_0=\frac{e^2 n\tau}{m_{\ast}},\,\,\,\,\,\,\omega_c=\frac{|e|B}{m_{\ast}c},
\end{equation}
where $B$ is the external magnetic field, $\tau$ is the transport scattering time due to non-magnetic impurities, $n$ is the sheet density of the mobile electrons, $e$ is an electron charge. 
Our model implies that the 
 magnetic field does not lead to the quantum oscillations, i.e. we assume that $\omega_c\tau<1$. 
THE can be described 
via effective spin-dependent cyclotron frequency $\Omega_s$:
\begin{equation}
\label{eq:Omegas}
\begin{aligned}
& \Omega_s = - n_{sk} v_s \int\limits_0^{2\pi} \Sigma_{ss}(\varphi) \sin{\varphi} d\varphi,
\end{aligned}
\end{equation}
 where $s$ denotes the spin projection of the scattered electron,
$v_s$ is the velocity at the Fermi level, which is spin-dependent due to the splitting of the spin subbands.
As we keep to the adiabatic regime, the asymmetric scattering rates have opposite sign for the opposite spin projections, so that $\Omega_{\uparrow} = - \Omega_{\downarrow} \equiv \Omega$. 
Thus, the topological contribution to the transverse conductivity is given by:
\begin{equation}
\label{eqsigmatop}
\sigma_{yx}^T=-\sigma_0\left(\Omega \tau\right)P_s,
\,\,\,\,\,\,\, P_s=\frac{n_\uparrow-n_\downarrow}{n_\uparrow+n_\downarrow}= \frac{\Delta}{2E_F},
\end{equation}
where $P_s$ is the spin polarization of the electrons in the system due to the background magnetization.
Eq.~(\ref{eqsigmatop}) is valid for small values of  $\Omega \tau$ when the 
asymmetric scattering on skyrmions is less effective than scattering on background impurities.

In Fig.~\ref{FigHall} the dependence of the total  transverse conductivity on the external magnetic field is shown together with 
the partial contributions from the  
ordinary $\sigma_{yx}^{O}$ and topological $\sigma_{yx}^{T}$ components.
These components are 
calculated according to the expressions (\ref{eqNormHall}) and (\ref{eqsigmatop}), respectively, with the same set of the parameters as in Fig.~\ref{Fig1}. The skyrmion profile $\Theta(r)$ and its size $a$ used in the calculations are the same 
as in Figs.~\ref{profile},\ref{FigRB}; Fig.~\ref{FigHall} was calculated for two values of the  skyrmion sheet density $n_{sk} = 2.5 \times 10^{11}$ cm$^{-2}$ and $n_{sk}= 1.25 \times 10^{11}$ cm$^{-2}$.


For the given parameters ($E_F =1$ eV, $\Delta=1$ eV, $m=0.6 m_0$) the topological contribution dominates over the ordinary Hall effect up to external magnetic field of $B \sim 2-4$ T.  
With the increase of the external magnetic field the size of the skyrmion is decreased leading to reduction of the topological contribution while the normal Hall conductivity linearly increases with $B$.
For the skyrmion configuration $m=1$ the sign of THE is opposite to the ordinary Hall effect, therefore 
when the magnitude of the two contributions become equal, the sign of the Hall effect is reversed. 
The situation is different for antiskyrmion, since for the negative vorticity $m=-1$ the sign of both contributions is the same. 
The value of the external magnetic field corresponding to the same magntitude of the two contributions 
depends on the skyrmions sheet density as shown in Fig.\ref{FigHall}. The sign change occurs at  $B\approx 4$ T for $n_{sk} = 2.5 \times 10^{11}$ cm$^{-2}$ (frame a), and $B\approx 2.5$ T for $n_{sk} = 1.25 \times 10^{11}$ cm$^{-2}$ (frame b). 

It is worth noting that the dependence of $\sigma_{yx}^T$ on the skyrmion size for the given set of parameters is rather weak, 
making THE quite robust in the wide range of the magnetic fields.  
It is due 
to the saturation of asymmetric scattering rates at large $k_F a$, which is typical for ferromagnets. 
The dependence of THE on $a$ is more pronounced for a lower Fermi energy (for example, due to stronger size quantization) so that $k_F a \le 20$ (Fig.~\ref{Fig1}).
Moreover, at $k_F a \le 10$ the spin-chirality driven contribution to the Hall effect arises~\cite{prl_skyrmion}, which leads to the crossover regime with a non-trivial oscillations of $\sigma_{yx}^T$~\cite{ourSciRep2017}. 

\section{Summary}
We have considered
topological Hall effect due to electron  scattering on individual magnetic skyrmions 
in the presence of external magnetic field.
At a relatively strong magnetic field ($B \gtrsim 1$ T), when the background magnetization is saturated and the anomalous Hall contribution does not depend on the magnetic field, 
THE manifests itself as an additional variation of Hall signal upon changing $B$. 
Our calculations show, that for thin films of strong ferromagnets 
the dependence of THE on the external magnetic field 
is rather weak. Although the skyrmion size changes in quite a wide range from $3$ to $12$ nm, the magnitude of THE does not exhibit the strong variation due to the saturation of the 
asymmetric scattering cross-section in the adiabatic regime.  
Therefore, the magnitude of THE is primarily determined by the skyrmions sheet density $n_{sk}$, 
 being comparable to the magnitude of the ordinary Hall effect (OHE) at $n_{sk}\sim 10^{11}$ cm$^{-2}$.
The two contributions have opposite sign for skyrmions with positive vorticity and, therefore partly compensate each other. For the array of antiskyrmions the signs are the same so the two contributions are added resulting in a larger Hall response. 




\section*{Acknowledgments}
The work
has been carried out under the financial support
from Russian Science Foundation,
project 17-12-01182 (analytical theory of asymmetric scattering) and
project 17-12-01265; (numerical calculations of topological Hall effect),
Russian Foundation of Basic Research (Project 18-02-00267) (calculation of skyrmion
configuration). M.N.P. and V.M.U. thank Icelandic scientific fund for the support. K.S.D. thank the Foundation for the Advancement of Theoretical Physics and Mathematics "BASIS".

\bibliography{Skyrmion-1}

\end{document}